\documentclass[aps,pre,reprint,groupedaddress,amsmath,amssymb,amsfonts,floatfix]{revtex4-2}

\usepackage{float}
\usepackage{graphicx}
\usepackage[caption=false]{subfig}
\usepackage[colorlinks,linkcolor=blue,anchorcolor=blue,citecolor=blue,urlcolor=blue]{hyperref}

\begin{document}


\title{Coevolutionary game dynamics with localized environmental resource feedback}


\author{Yi-Duo Chen}

\author{Jian-Yue Guan}
\email[Corresponding author: ]{guanjy@lzu.edu.cn}

\author{Zhi-Xi Wu}

\affiliation{Lanzhou Center for Theoretical Physics, Key Laboratory of Theoretical Physics of Gansu Province, and Key Laboratory of Quantum Theory and Applications of MoE, Lanzhou University, Lanzhou, Gansu 730000, China}%
\affiliation{Institute of Computational Physics and Complex Systems, Lanzhou University, Lanzhou, Gansu 730000, China}


\date{\today}

\begin{abstract}
    Dynamic environments shape diverse dynamics in evolutionary game systems. 
    We introduce spatial heterogeneity of resources into the prisoner's dilemma game model to explore coevolutionary game dynamics with environmental feedback. 
    The availability of resources significantly affects the survival competitiveness of surrounding individuals. 
    Feedback between individuals' strategies and the resources they can use leads to the oscillating dynamic known as the ``oscillatory tragedy of the commons''.  
    Our findings indicate that when the influence of individuals' strategies on the update rate of resources is significantly high in systems characterized by environmental heterogeneity, they can attain an equilibrium state that avoids the oscillatory tragedy. 
    In contrast to the numerical results obtained in well-mixed structures, self-organized clustered patterns emerge in simulations utilizing square lattices, further enhancing the stability of the system. 
    We discuss critical phenomena in detail, demonstrating that the aforementioned transition is robust across various system parameters, including the strength of cooperators in restoring the environment, initial distributions of cooperators, system size and structures, and noise. 
    
\end{abstract}


\maketitle


\section{Introduction}

Coevolutionary game dynamics with environmental feedback have become topics of significant interest in complex systems science in recent years \cite{weitz2016oscillating,tilman2020Evolutionary,hauert2019asymmetric,hilbe2018Evolution,su2019Evolutionary,kleshnina2023effect,huang2020learning,wang2021evolution,lin2019Spatial}.  
In an ecosystem with limited public goods, individuals will inevitably defect in pursuit of their own benefit, ultimately depleting available resources. 
Hardin referred to this dilemma as ``the tragedy of the commons'' (TOC), asserting that there is no technical solution to it \cite{hardin1968tragedy}. 
Recently, different authors have studied dynamics related to this TOC using replicator dynamics \cite{hofbauer1998Evolutionary,traulsen2005Coevolutionary,traulsen2006Coevolutionary,traulsen2006Stochastic}. 
Weitz \textit{et al.}~studied a class of dynamics known as the ``oscillatory tragedy of the commons'' (o-TOC) which incorporates payoffs with environmental feedback \cite{weitz2016oscillating}. 
In this model, defectors degrade the environment while cooperators restore it. 
Thus, the system cycles between deplete and replete environmental states, and between cooperation and defection behavior states. 
Individuals tend to choose cooperation in the depleted state and defection in the replete state respectively to achieve higher payoffs. 
For instance, overgrazing degrades grasslands, which can be restored through controlled grazing \cite{weitz2016oscillating}. 
Tilman \textit{et al.}~proposed a comprehensive framework for ecoevolutionary games, providing examples and analyzing the dynamic mechanisms underlying the oscillation phenomena \cite{tilman2020Evolutionary}. 
Hauert \textit{et al.}~investigated asymmetric evolutionary games in heterogeneous environments consisting of patches of different qualities and discovered a range of dynamic phenomena \cite{hauert2019asymmetric}. 
However, the aforementioned studies of replicator dynamics \cite{weitz2016oscillating,tilman2020Evolutionary,hauert2019asymmetric} assumed homogenized structures and environmental feedback, which may not accurately reflect real-world systems.

To address these challenges, some authors have considered heterogeneous interactions between individuals in Monte Carlo (MC) simulations, which can uniquely impact the dynamics of the system. 
Hilbe \textit{et al.}~and Su \textit{et al.}~found that game transitions can effectively promote cooperation rates \cite{hilbe2018Evolution,su2019Evolutionary}, representing the effect of localized environmental feedback in evolutionary games and relating to the idea of partner-fidelity feedback in evolutionary biology, such as grass-endophyte mutualism \cite{bull1991distinguishing,sachs2004evolution,schardl1997evolution,cheplick2009ecology}. 
Kleshnina \textit{et al.}, Huang \textit{et al.}~and Wang \textit{et al.}~have also utilized game transitions to investigate the effects of game information, reinforcement-learning individuals and state-dependent strategies, respectively \cite{kleshnina2023effect,huang2020learning,wang2021evolution}. 
However, research related to game transitions has generally focused on the evolution of cooperation, lacking attention to the coevolutionary dynamics of the environment. 
Studies of ecosystems, such as mussel beds, arid ecosystems, and bacterial colonies, have considered spatial heterogeneity of the environment \cite{liu2014pattern,kefi2008evolution,kummerli2010molecular}. 
Additionally, Halatek \textit{et al.}~found that, compared with nonspatial models, spatial interactions can significantly affect system stability and promote the formation of spatial patterns \cite{halatek2018Rethinkinga}. 
In summary, heterogeneous environmental feedback has significant influences on evolutionary dynamics that should not be ignored.

To address these issues, Lin \textit{et al.}~considered spatiotemporal dynamics with environmental diffusion in an individual-based model and analyzed the dynamic mechanism of o-TOC \cite{lin2019Spatial}. 
They assigned environmental resources to each node to characterize the heterogeneous distribution of environments in real systems with heterogeneous feedback on interactions between individuals. 
They discovered that spatial interactions can significantly influence the parameter domains, such as payoff matrices, where o-TOC occurs, compared with nonspatial models. 
Thus, in this work, we introduce spatially heterogeneous resources into the prisoner's dilemma game model with environmental feedback. 
Unlike previous studies, environmental resources are assigned to nodes in square lattices, which can be used by individuals on adjacent nodes and cannot spread around. 
Resources affect the payoffs of the game for surrounding individuals and are subject to change by the actions of neighboring individuals. 
To capture the dynamics of individuals who are sensitive to environmental changes and update their strategies immediately, as illustrated by the aforementioned example of overgrazing, we employ myopic rules \cite{sysi2005spatial, chen2009cooperation, roca2009promotion} of strategy update in our model. 
o-TOC dynamics are also found in our results. 
Additionally, when the update rate of resources is sufficiently high, a transition from the oscillating state to an equilibrium state occurs. 
By comparing the results of theoretical approximations, we demonstrate that environmental resource heterogeneity and the structure of inter-individual interactions significantly affect the stability of coevolutionary game systems. 

The remainder of the article is organized as follows. 
In Sec.~\ref{model}, we present our model, detail the MC simulations, and theoretically analyze the temporal evolution of the coevolutionary game dynamics. 
In Sec.~\ref{rd_a}, we explore the phenomena and causes underlying the oscillating dynamics. 
Building upon this foundation, we identify the transition from the oscillating state to the equilibrium state and discuss critical values and phenomena in detail in Sec.~\ref{rd_b}. 
Finally, in Sec.~\ref{conclusion}, we summarize our work and findings.

\section{\label{model}Model and simulations}

\subsection{\label{ibm}Individual-based coevolutionary game dynamics}

We study the evolutionary dynamics of $2 \times 2$ games in square lattices with $N=L \times L$ nodes and periodic boundary conditions. 
Every individual $i$ takes up a node alone and has a strategy $s_i \in \{ C, D \} $, $i=1, 2, ...,N$. 
$C$ stands for cooperation and $D$ stands for defection. 
Initially, individuals' strategies are randomly assigned as cooperation with equal probability $x_{C0}$ or defection with probability $1-x_{C0}$. 
Each individual $i$ is assigned a certain amount of resources, represented by the parameter $e_i \in [0,1]$.  
$e_i=0$ means node $i$ has the poorest resources and $e_i=1$ means node $i$ has the most abundant resources. 
Individuals can use their own and their neighbors' resources. 
Thus, the total resources individual $i$ can use are 
\begin{equation}\label{e1}
    n_i=0.5\times e_i + 0.125\times(e_a + e_b + e_c + e_d), 
\end{equation}
$a, b, c, d$ are the neighbors of individual $i$, and $e_a, e_b, e_c, e_d$ are their resources respectively. 
When individual $i$ plays games with their neighbors, their payoffs are determined by $n_i$ and the following payoff matrices:
\begin{equation}\label{e2}
    \Pi_i=(1-n_i)\times A_0 + n_i\times A_1, 
\end{equation}
where $A_0$ and $A_1$ are the matrices that correspond to situations of resource scarcity and resource abundance respectively. 
$A_1$ represents the payoff matrix of the prisoner's dilemma game which is used when resources are abundant and $A_0$ corresponds to the payoff matrix that benefits cooperators when individuals deplete their resources (i.e., $A_0$ is not a payoff matrix of the Prisoner's Dilemma Game), respectively. 
The advantage of cooperators in $A_0$ arises because cooperators can regenerate resources in hostile environments. 
\begin{equation}\label{e3}
    A_0=
    \begin{bmatrix}
        R_0 & S_0 \\
        T_0 & P_0 \\
    \end{bmatrix}, \quad
    A_1=
    \begin{bmatrix}
        R_1 & S_1 \\
        T_1 & P_1 \\
    \end{bmatrix},
\end{equation}
where $R_0$($R_1$) denotes the reward for cooperation, $S_0$($S_1$) denotes the sucker's payoff, $T_0$($T_1$) denotes the temptation to cheat, and $P_0$($P_1$) denotes the punishment for mutual defection. 

At each step $\tau$, we randomly choose an individual $i$ to play games with its neighbors and update its strategy $s_i$ to $s_{i'}$ by myopic rules \cite{sysi2005spatial,chen2009cooperation,roca2009promotion} with probability 
\begin{equation}\label{e4}
    P(s_{i'} \to s_i)=\frac{1}{1+{\rm exp}(\frac{\pi_i-\pi_{i'}}{k})} , 
\end{equation}
where $\pi_i$ is the average payoff of individual $i$'s games with four neighbors, $s_{i'}$ is the opposed strategy of $s_i$, $\pi_{i'}$ is the average payoff of individual $i$ with strategy $s_{i'}$, and $k$ controls the noise. 
After updating individual $i$'s strategy $s_i(\tau)$ to $s_i(\tau +1)$, we update node $i$'s and its neighbors' resources
\begin{equation}\label{e5}
    e_j(\tau +1)=
    \begin{cases}
        e_j(\tau)+\alpha \mu \theta \qquad s_i(\tau +1)=C, \\
        e_j(\tau)-\alpha \mu \qquad \enspace s_i(\tau +1)=D, \\
    \end{cases}
\end{equation}
where $j=i, a, b, c, d$. 
$\mu \in [0, 1]$ is the parameter that controls the update rate of resources relative to the update rate of strategies; $\theta \in (0, 1/\mu]$ denotes the strength of cooperators in restoring the environment;$\alpha = 0.5$ when $j=i$; and $\alpha = 0.125$ when $j=a, b, c, d$. 
We restrict $e_i \in [0,1]$ while updating resources. 
If $e_i$ should be updated to a value greater (less) than 1 (0) according to Eq.~\eqref{e5}, we set it to 1 (0). 
We take $N$ steps with chances to update strategies and resources as an MC step, and every individual has an equal chance to update its strategy once in each MC step.

The fraction of cooperators is defined as 
\begin{equation}\label{rxc}
    x_C=\frac{1}{N}\sum_{i=1}^{N}\delta(s_i-C),
\end{equation}
where $\delta(s_i-C)=1$ when $s_i=C$ and $\delta(s_i-C)=0$ when $s_i=D$. 
The average resources of the system is 
\begin{equation}\label{rea}
    e=\frac{1}{N}\sum_{i=1}^{N}e_i. 
\end{equation}
We use these two parameters to discuss the coevolution of the cooperation level and the average resources of the system. 
In our simulations, without loss of generality, we set noise $k=0.01$ and system size $N=L\times L=400\times 400$. 
The payoff matrices are
\begin{equation}\label{rpf}
    A_0=
    \begin{bmatrix}
        \enspace5 &\enspace 2\enspace \\
        \enspace4 &\enspace 2\enspace \\
    \end{bmatrix}, \quad
    A_1=
    \begin{bmatrix}
        \enspace5 &\enspace 0\enspace \\
        \enspace6 &\enspace 2\enspace \\
    \end{bmatrix}, 
\end{equation}
where $A_1$ is the payoff matrix of the prisoner's dilemma game, and $S_0 > S_1$ and $T_0 < T_1$ are set to benefit cooperators in $A_0$, which represents a hostile environmental situation. 
$R_0 = R_1$ and $P_0 = P_1$ are set to examine a simplified scenario in which mutual cooperation (or defection) consistently plays the same role, regardless of whether the situation involves resource scarcity ($A_0$) or resource abundance ($A_1$). 
In our MC simulation results, the initial probability for individuals to cooperate is $x_{C0}=0.2$ and the initial resource of nodes are all set to $e_0=0.3$. 
It is important to note that the values of $x_{C0}$ and $e_0$ have no significant influence on our results and conclusions. 
The system stabilized in all the results presented, either by reaching an oscillating or equilibrium state, and extending the simulation time does not lead to any new phenomena.

\subsection{\label{mft}Theoretical analysis}

In addition to the MC simulations just described, we employ mean-field theory to investigate the temporal evolution of the fraction of cooperators and the average resources of the system. 
Individuals in the system are well-mixed and interact with four randomly chosen individuals, based on the structure of the square lattice used in the MC simulations. 
To characterize the effects of environmental heterogeneity, we assume there are only two states of environmental resources in theoretical analysis: an abundant state (i.e., $e_i=1$) and an infertile state (i.e., $e_i=0$). 
In this case, we define the probability of a node with abundant resources as $P_{a, i}=e=\frac{1}{N}\sum_{i}e_i$, where $e_i$ represents the environmental resources of node $i$ and only utilized in the MC simulations. 

Thus, when the average environmental resources of system is $e \in [0, 1]$, the probability $P_e(h)$ that an individual has $h \in \{0, 1, 2, 3, 4\}$ nodes with abundant resources among its neighbors is defined as follows:
\begin{equation}\label{ab1}
    P_e(h)=\binom{4}{h}e^{h}(1-e)^{4-h}. 
\end{equation}
Moreover, when the fraction of cooperation is $x_C \in [0, 1]$, the probability $P_{x_C}(g)$ that an individual has $g \in \{0, 1, 2, 3, 4\}$ cooperative neighbors is given by:
\begin{equation}\label{ab2}
    P_{x_C}(g)=\binom{4}{g}x_C^{g}(1-x_C)^{4-g}. 
\end{equation}
Based on Eq.~\eqref{e1}, an individual's usable resources $n(e, h)$ is determined by its own and its neighbors' nodes' resources:
\begin{equation}\label{ab3}
    \begin{aligned}
        n(e, h)&=0.5(e\times 1+(1-e)\times 0)\\
        &\enspace+0.125(h\times 1+(4-h)\times 0)\\
        &=e/2+h/8.  
    \end{aligned}
\end{equation}
According to Eq.~\eqref{e5}, node $i$ updates its resources at each MC step based on its own and its neighbors' strategies: 
\begin{equation}\label{ab4}
    \begin{aligned}
        e_i(t+1)&=e_i(t)+0.5(\theta x_C^{'}-(1-x_C^{'}))\\
        &\enspace+0.125(\theta g-(4-g)),
    \end{aligned}
\end{equation}
where $x_C^{'}$ is individual $i$'s probability of cooperation (details of the calculations are provided in Appendix). 
Based on Eqs.~\eqref{aa7} and \eqref{ab3}, the probability of cooperation for each individual $i$ is expressed as: 
\begin{equation}\label{ab5}
    x_C^{'}=\frac{1}{1+{\rm exp}((2n(e, h)-0.25g)/k)}. 
\end{equation}

Using Eqs.~\eqref{ab1}\textendash \eqref{ab5}, the temporal evolution of the fraction of cooperators $x_C$ is described by:
\begin{equation}\label{abx}
    \begin{aligned}
        \dot{x_C}&=\sum_{h=0}^{4}\sum_{g=0}^{4}P_e(h)P_{x_C}(g)(-x_C(1-x_C^{'})\\
        &\enspace +(1-x_C)x_C^{'})\\
        &=\sum_{h=0}^{4}\sum_{g=0}^{4}\binom{4}{h}\binom{4}{g}e^{h}(1-e)^{4-h}x_C^{g}(1-x_C)^{4-g}\\
        &\enspace \times (\frac{-x_C}{1+{\rm exp}((-2n(e, h)+0.25g)/k)}\\
        &\enspace +\frac{1-x_C}{1+{\rm exp}((2n(e, h)-0.25g)/k)}), 
    \end{aligned}
\end{equation}
and the temporal evolution of the average environmental resources $e$ is given by:
\begin{equation}\label{abe}
    \begin{aligned}
        \dot{e}&=\sum_{h=0}^{4}\sum_{g=0}^{4}P_e(h)P_{x_C}(g)(0.5(\theta x_C^{'}-(1-x_C^{'}))\\
        &\enspace+0.125(\theta g-(4-g)))\\
        &=\sum_{h=0}^{4}\sum_{g=0}^{4}\binom{4}{h}\binom{4}{g}e^{h}(1-e)^{4-h}x_C^{g}(1-x_C)^{4-g}\\
        &\enspace \times (\frac{-0.5}{1+{\rm exp}((-2n(e, h)+0.25g)/k)}\\
        &\enspace +\frac{0.5\theta}{1+{\rm exp}((2n(e, h)-0.25g)/k)}+\frac{g\theta-(4-g)}{8}). 
    \end{aligned}
\end{equation}
By numerically integrating Eqs.~\eqref{abx} and \eqref{abe} using the fourth-order Runge-Kutta method (we refer to these later as the numerical results), we can obtain the coevolutionary dynamics of the fraction of cooperators and the average resources of the system, and the transition from the oscillating state to the equilibrium state. 
Initial values of $x_{C0} \in (0, 1)$ and $e_0 \in (0, 1)$ also have no significant influence on numerical results; we choose $x_{C0}=0.2$ and $e_0=0.3$. 

Overall, compared with the MC simulations, the numerical results obtained using Eqs.~\eqref{abx} and \eqref{abe} provide a less detailed representation of the heterogeneous distribution of environmental resources and do not account for fixed neighbors among individuals, which may lead to stable distributions.

\section{\label{result}Results and discussion}

\subsection{Oscillating dynamics\label{rd_a}}

\begin{figure*}
    \includegraphics[width=\linewidth]{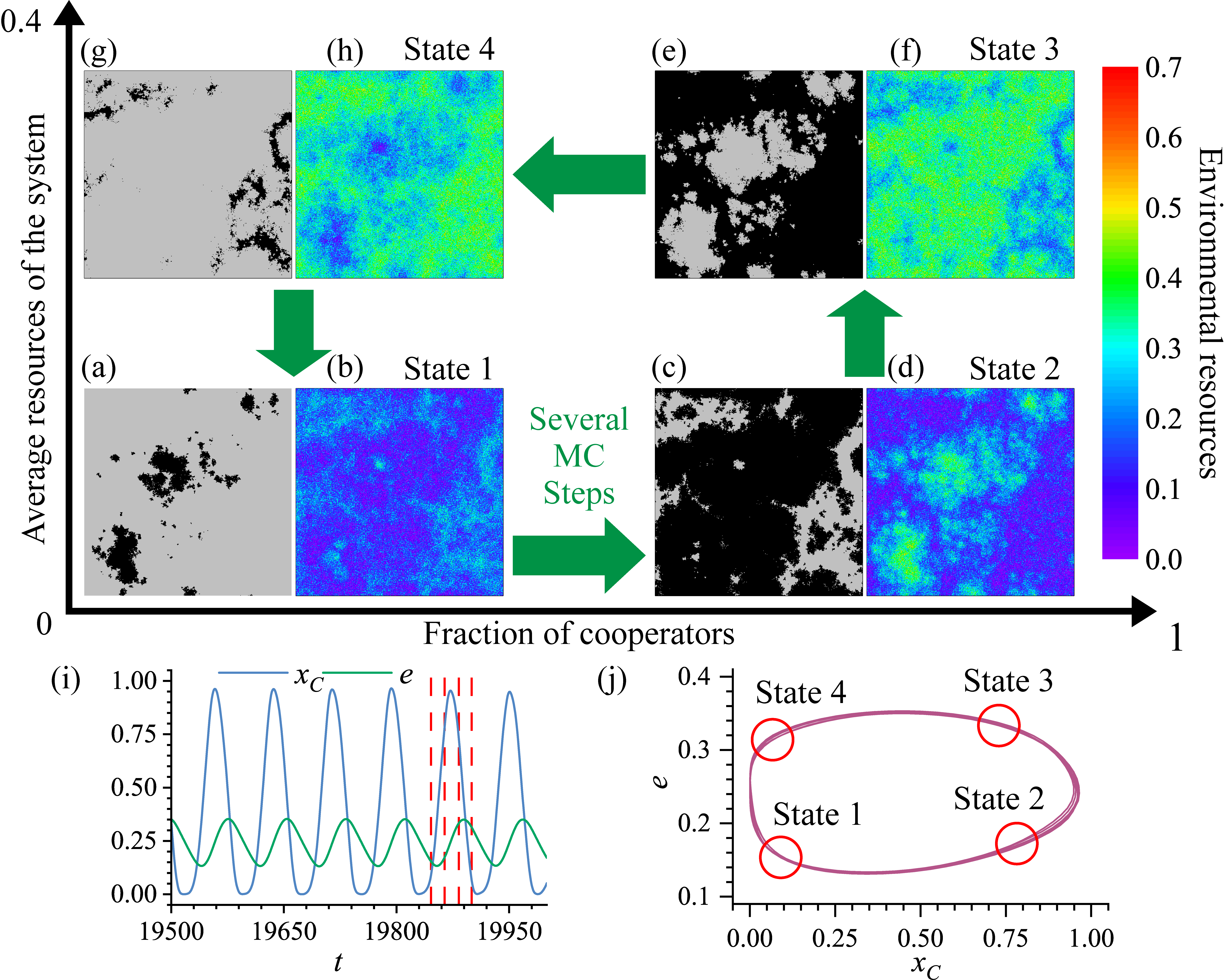}

  \caption{\label{fig1}Four states of the oscillating coevolutionary dynamics of strategies and resources in the phase diagram of simulation results, represented by the distributions of the system for $\theta=1.5$ and $\mu=0.007$ at (a), (b) $t=19 \thinspace 846$; (c), (d) $t=19\thinspace 864$; (e), (f) $t=19\thinspace 883$; and (g), (h) $t=19\thinspace 900$. (a), (c), (e), (g) Distributions of cooperators (black pixels) and defectors (gray pixels). (b), (d), (f), (h) Distributions of environmental resources of nodes in the system. (i) Time evolution of the fraction of cooperators and the average resources of the system from $t=19\thinspace 500$ to $t=20\thinspace 000$. The aforementioned four states are marked by four vertical red dashed lines in order from left to right. (j) Phase diagram of the coevolutionary dynamics of strategies and resources from $t=19\thinspace 500$ to $t=20\thinspace 000$. The regions of these four states are marked with red circles. 
  }    
  
\end{figure*}

\begin{figure*}
    \includegraphics[width=\linewidth]{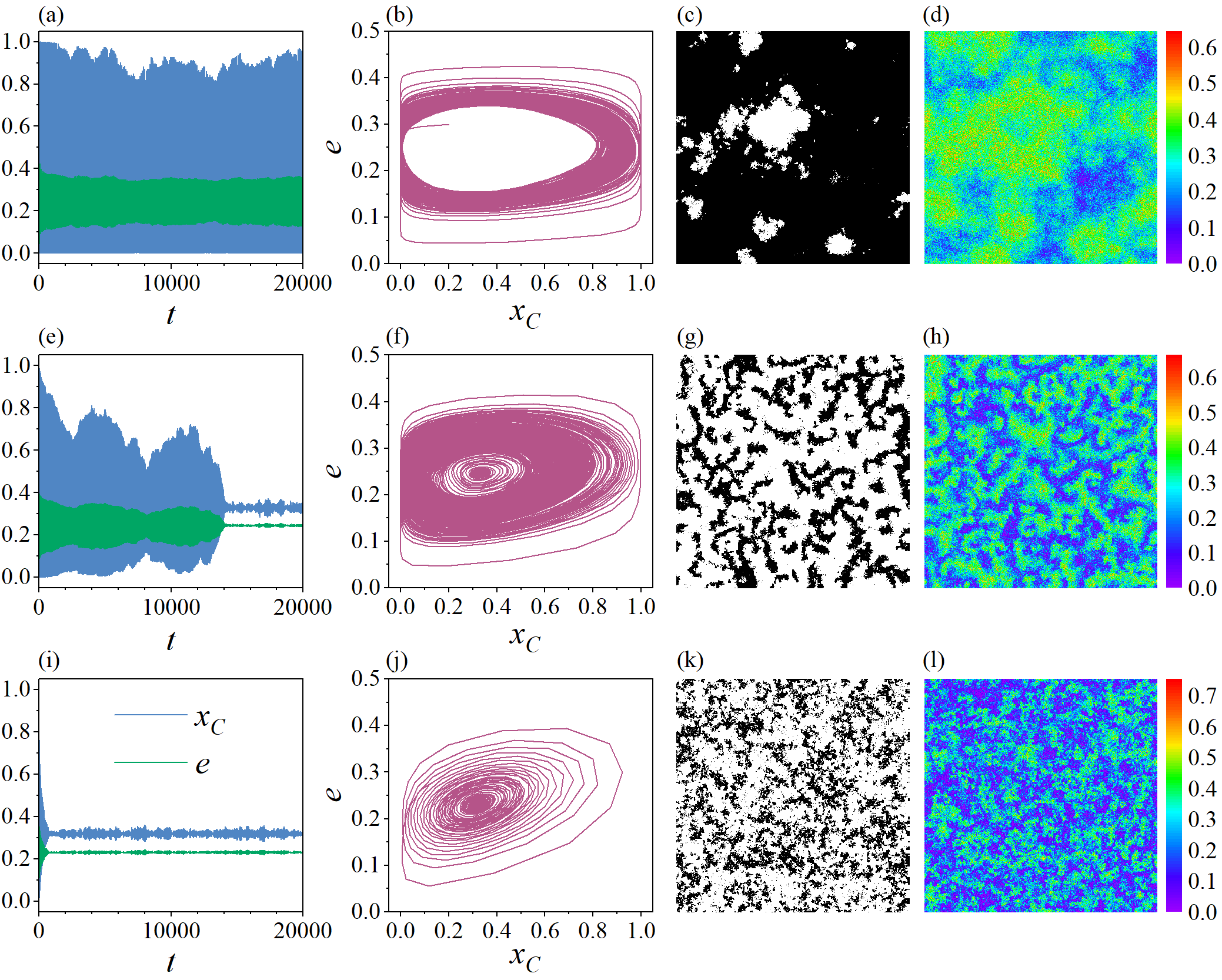}

  \caption{\label{fig2}Simulation results with different evolutionary dynamics for $\theta=2.0$ and different update rate of resources (a)\textendash (d) $\mu=0.005$, (e)\textendash (h) $\mu=0.016$, and (i)\textendash (l) $\mu=0.04$, respectively. (a), (e), (i) Time evolutions of the fraction of cooperators and the average resources of the system. (b), (f), (j) Phase diagram of the coevolutionary dynamics of strategies and resources. (c), (g), (k) Distributions of cooperators (black pixels) and defectors (white pixels). (d), (h), (l) Distributions of environmental resources of nodes in the system. }
    
\end{figure*}

\begin{figure*}

    \includegraphics[width=\linewidth]{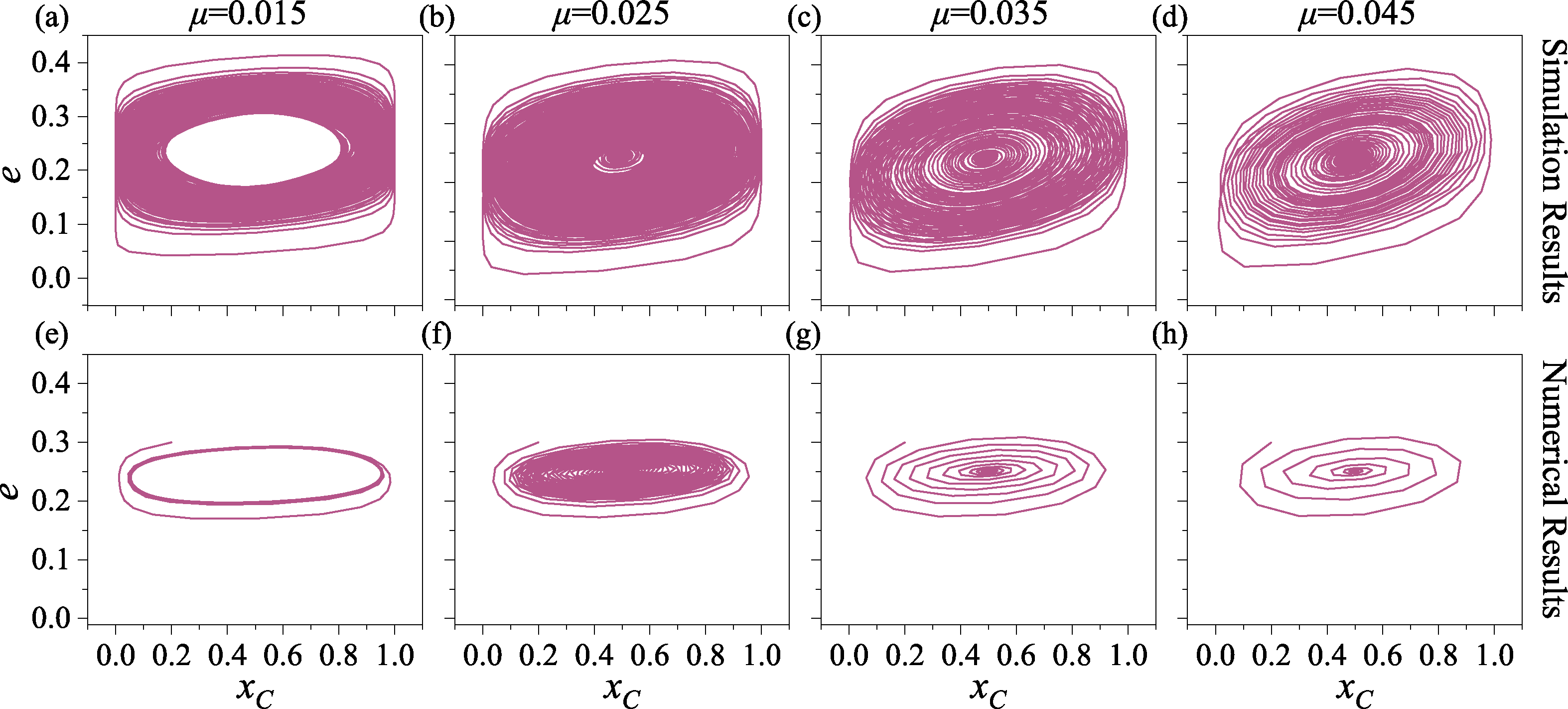}

  \caption{\label{fig3}Phase diagram of the coevolutionary dynamics of strategies and resources of (a)\textendash (d) simulation results and (e)\textendash (f) numerical results for $\theta=1.0$ and different values of $\mu$. (a), (e) $\mu=0.015$; (b), (f) $\mu=0.025$; (c), (g) $\mu=0.035$; and (d), (h) $\mu=0.045$. } 
    
\end{figure*}

From Fig.~\ref{fig1} we can see that the system exhibits persistent oscillating dynamics of the fraction of cooperators $x_C$ and the average resources of the system $e$. 
Upon reaching a periodic oscillating state, the system cycles through four distinct states within each period: (1) deficient resources and low fraction of cooperators ($x_{C,\mathrm{min}}$, $e_{\mathrm{min}}$), (2) deficient resources and high fraction of cooperators ($x_{C,\mathrm{max}}$, $e_{\mathrm{min}}$), (3) sufficient resources and high fraction of cooperators ($x_{C,\mathrm{max}}$, $e_{\mathrm{max}}$), and (4) sufficient resources and low fraction of cooperators ($x_{C,\mathrm{min}}$, $e_{\mathrm{max}}$). 

Assume the system initially starts in state 1, characterized by deficient environmental resources and dominance of defectors.
Individuals gradually choose cooperation to obtain higher payoffs in the barren environment, the environmental resources remain at a low level during this process, and the system evolves to state 2. 
At this point, the dominant cooperators gradually restore sufficient environmental resources, leading the system to evolve to state 3, where cooperators maintain their dominant position. 
Moreover, individuals tend to choose defection for higher payoffs, and defectors quickly occupy almost the entire system before the environmental resources are significantly depleted, reaching state 4. 
After this, the environmental resources are gradually consumed by the defectors, returning the system to state 1. 

The distributions of the four states in Fig.~\ref{fig1} are taken from consecutive periods of time after the simulation has stabilized. 
They are marked by four vertical red dashed lines in the time series of Fig.~\ref{fig1}(i) in order from left to right. 

Figures \ref{fig2}(a) and (b) illustrate the time evolution of the fraction of cooperators and the average resources of the system and the phase diagram of the coevolutionary dynamics of strategies and resources, respectively. 
In Fig.~\ref{fig2}(b), the phase trajectory exhibits periodic rotations within a region similar to an irregular rounded quadrilateral, the four vertices regions of the quadrilateral correspond to the four states previously described [they are also shown more clearly in Fig.~\ref{fig1}(j)]. 
Oscillating dynamics are found in both simulation and numerical results for similar values of low $\mu$ [see Figs.~\ref{fig3}(a) and \ref{fig3}(e)]. 
Distributions at $t=20\thinspace 000$ are shown in Figs.~\ref{fig2}(c) and (d); clusters of cooperators and defectors appear, expand, and disappear rapidly and do not last.

According to the myopic rule [see Eq.~\eqref{e4} and Appendix], the probability of cooperation is not influenced by individuals' previous strategies. 
Based on Eqs.~\eqref{e1}\textendash \eqref{e4}, given the neighbors' strategies and the total amount of resources available to the individuals, the individuals' probability of cooperation can be calculated, see details in Appendix. 
Results in Fig.~\ref{figa1} further explain the aforementioned mechanism of oscillations, cooperators tend to defect when resources are sufficient or when there are fewer cooperative neighbors, while defectors are inclined to cooperate when resources are deficient or when more neighbors are cooperative. 
Additionally, the interplay between cooperators' production and defectors' consumption of resources drives the system through these cyclical states. 
This oscillatory behavior is analogous to the o-TOC dynamics described in work of Weitz \textit{et al.}~\cite{weitz2016oscillating}.

\subsection{\label{rd_b}Transition to the equilibrium state}

\begin{figure*}

    \includegraphics[width=\linewidth]{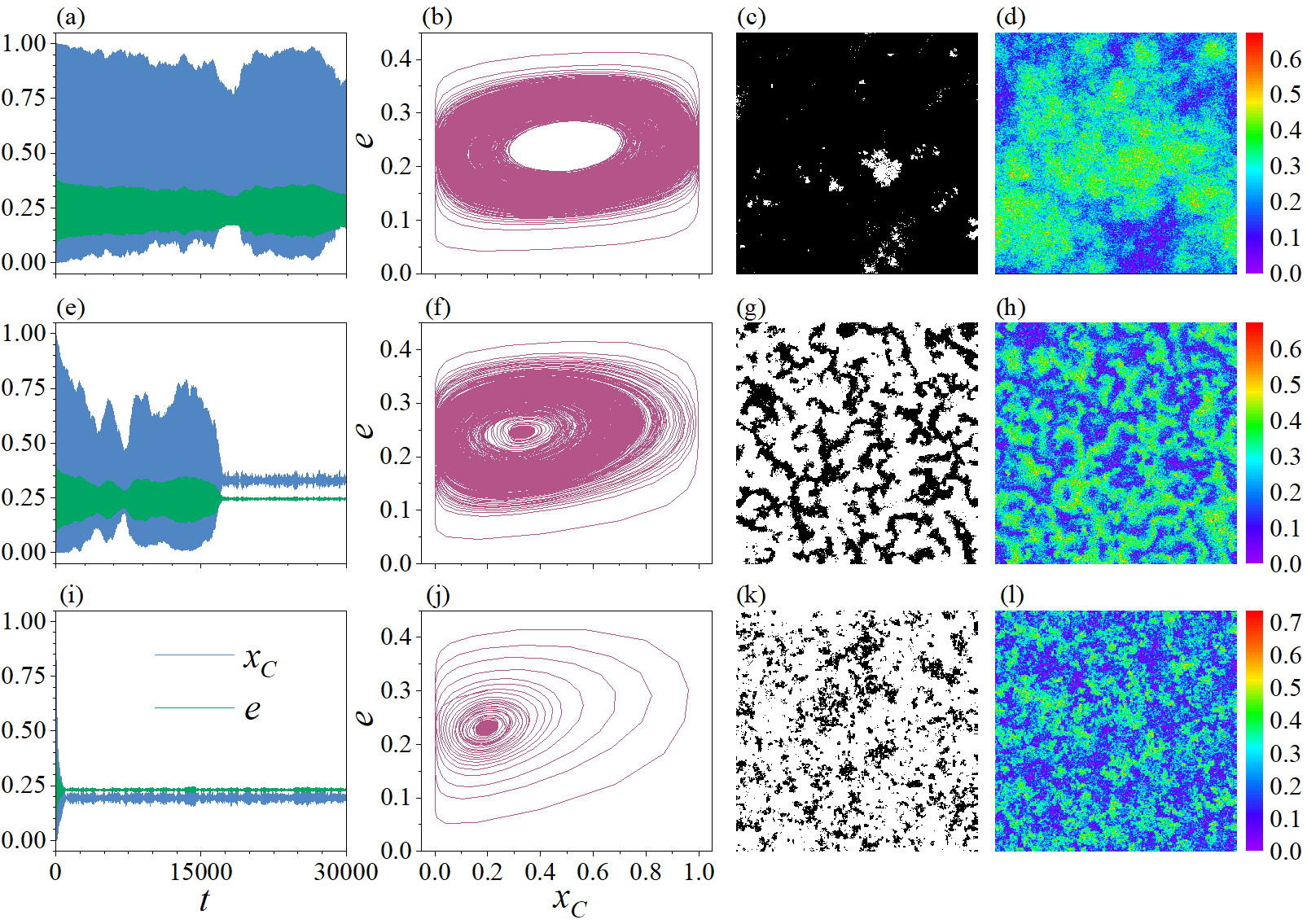}

  \caption{\label{fig4}Simulation results with different evolutionary dynamics for $\mu=0.015$ and different strength of cooperators in restoring the environment (a)\textendash (d) $\theta=1.0$, (e)\textendash (h) $\theta=2.0$, and (i)\textendash (l) $\theta=4.0$, respectively. (a), (e), (i) Time evolutions of the fraction of cooperators and the average resources of the system. (b), (f), (j) Phase diagram of the coevolutionary dynamics of strategies and resources. (c), (g), (k) Distributions of cooperators (black pixels) and defectors (white pixels). (d), (h), (l) Distributions of environmental resources of nodes in the system. } 
    
\end{figure*}

When we keep the value of $\theta$ constant, a significant increase in the update rate of resources $\mu$ results in the disappearance of the previously observed persistent oscillations [see Figs.~\ref{fig2}(a), \ref{fig2}(b), \ref{fig2}(e) and \ref{fig2}(f)]. 
As shown in Figs.~\ref{fig2}(e) and \ref{fig2}(f), after several periods, the system transitions from an oscillating transient state to an equilibrium state. 
In the time series, the amplitudes of the fluctuations rapidly decrease within a short period, eventually resulting in only minor fluctuations [see Fig.~\ref{fig2}(e)]. 
As shown in Fig.~\ref{fig2}(f), following several periodic rotations, the phase trajectory gradually converges toward the center. 
Importantly, the duration of the oscillating transient state before reaching equilibrium in simulations is unstable across simulations. 
Even with identical parameters, different systems reach equilibrium at varying rates.
Figures \ref{fig2}(g) and (h) reveal clustered spatial patterns of cooperators, defectors, and resources, which emerge as the system transitions to the equilibrium state. 
We demonstrate that these clustered patterns emerge during the transition stage to the equilibrium state. 
The equilibrium state also appears in the numerical results for greater value of $\mu$ [see Figs.~\ref{fig3}(a) and (b)]. 
However, for the same value of $\mu$ and $\theta$, the numerical results and the simulation results do not display exactly the same phenomenon, and we will make a detailed comparison later. 

To further investigate the relations among parameters, oscillations, and patterns, we conducted several groups of simulations for varying values of $\mu$ and $\theta$, some of which are shown in Figs.~\ref{fig2} and \ref{fig4}. 
When $\mu$ is sufficiently low, persistent oscillations occur without sustained spatial patterns [Figs.~\ref{fig2}(a)\textendash (d)]. 
Conversely, as $\mu$ increases, clustered patterns emerge and the system eventually reaches an equilibrium state [see Figs.~\ref{fig2}(e)\textendash (h)]. 
Further increases in $\mu$ result in smaller cluster sizes and less time to enter the equilibrium state [see Figs.~\ref{fig2}(i)\textendash (l)]. 
By altering $\theta$, we observe different phenomena, including variations in the sizes of clustered patterns as the systems enter the equilibrium state [see Figs.~\ref{fig4}(e)\textendash (l)], and even the absence of sustained spatial patterns in the oscillating state [see Figs.~\ref{fig4}(a)\textendash (d)]. 

In summary, increasing the update rate of resources $\mu$ triggers a transition from the oscillating state to an equilibrium state. 
And the critical value $\mu_C$ is mainly influenced by $\theta$ (see Fig.~\ref{fig6}). 
However, even with identical parameters, certain systems may persist in the oscillating state without forming stable patterns when $\mu$ is near the critical value $\mu_C$, even after a long period of time. 
In our simulations, oscillations did not significantly influence the average fraction of cooperators, given the same other parameters. 

In systems characterized by large clusters, clusters gradually emerge as the amplitude of the time series begins to decrease. 
Additionally, spiral patterns appear at the beginning of the formation process of clustered patterns. 
Over time, these spirals eventually couple and form clustered patterns, and the system enter into an equilibrium state characterized by only slight fluctuations in the time series, with the movements and deformations of these clusters. 
In Ref.~\cite{patternformation}, we provide a GIF file showing the distributions of cooperators, defectors, and resources throughout the evolution of the system with the same parameters as those used in Figs.~\ref{fig2}(e)\textendash (h). 
The GIF file mainly shows the stage of the spiral patterns' emergence and their evolution into clustered patterns. 

As previously mentioned, the clustered patterns are not entirely static when systems enter the equilibrium state. 
Clusters of cooperators tend to migrate toward nodes with deficient resources, while clusters of defectors shift to nodes with sufficient resources (this phenomenon aligns well with results presented in Appendix and Fig.~\ref{figa1}). 
Moreover, cooperators produce resources in deficient nodes, whereas defectors consume resources in sufficient nodes. 
This local feedback dynamic results in a dynamic equilibrium throughout the system. 
As clusters of cooperators and defectors shift, the overall density of cooperators remains constant, thereby maintaining a stable average level of resources in the system.

Our findings also indicate that systems with a greater value of $\mu$ reach equilibrium faster and exhibit smaller clusters (see Figs.~\ref{fig2}(e)\textendash  \ref{fig2}(l) and \ref{fig4}(e)\textendash  \ref{fig4}(l)). 
A greater value of $\mu$ accelerates the evolution of environmental resources, which in turn speeds up the coevolution dynamics of strategies and resources on nodes, leading to a quicker attainment of equilibrium. 
In this case, cooperators require less time to restore resources, while defectors deplete them more swiftly. 
Furthermore, the local feedback dynamics of clusters occur only in more confined areas for higher values of $\mu$, where individuals can rapidly alter the environment, resulting in smaller clusters within the system. 
The spiral patterns during the transition from the oscillating state to the equilibrium state are less pronounced in systems with smaller clusters (i.e., systems with higher values of $\mu$ or different values of $\theta$). 
These spiral patterns are small and evolve rapidly, making them difficult to discern, which contrasts with the results in Ref.~\cite{patternformation}. 

It is important to note that the transition described above can only occur in systems with sufficiently large spatial scales (i.e., a sufficient number of nodes). 
Additionally, the critical value of $\mu$ is influenced by system size $N$. 
The conclusions drawn remain robust despite variations in parameters, including the strength of cooperators in restoring the environment and the initial distributions of cooperators. 
Furthermore, we demonstrate that these conclusions remains robust when the noise does not significantly affect the system (i.e., around the value $k=0.01$ used in this work).

\begin{figure*}

    \includegraphics[width=\linewidth]{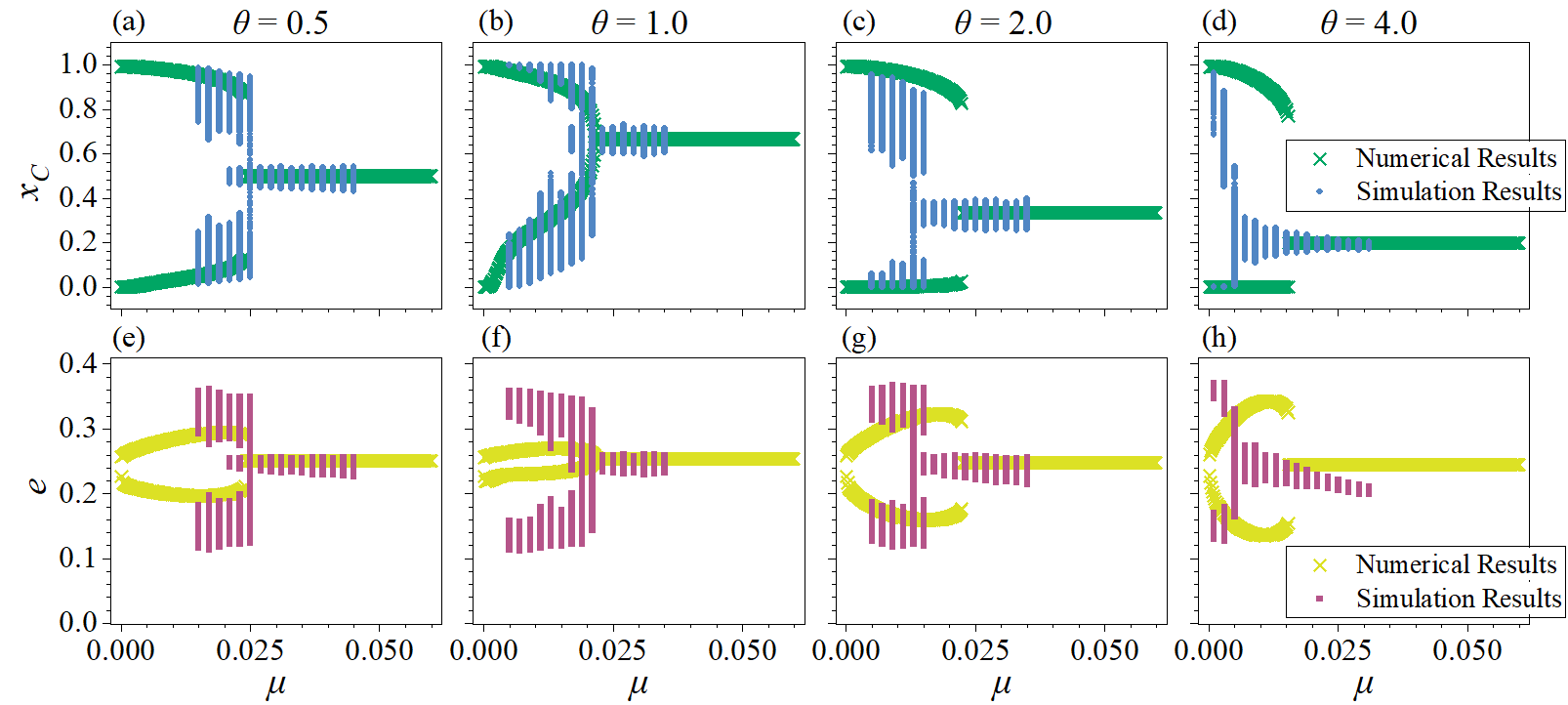}

  \caption{\label{fig5}Fraction of cooperators (a)\textendash (d) and average resources of the system (e)\textendash (h) as functions of $\theta$ of simulation and numerical results. (a), (e) $\theta=0.5$; (b), (f) $\theta=1.0$; (c), (g) $\theta=2.0$; and (d), (h) $\theta=4.0$. To accurately describe the system's oscillating dynamics, we record the local maxima and minima of both the fraction of cooperators and the average resources of the system in the Monte Carlo simulations from $t=35\thinspace 000$ to $t=40\thinspace 000$. And in the numerical results, we capture the local maxima and minima during the oscillating state, as well as the stabilized values in the equilibrium state. } 
    
\end{figure*}

\begin{figure}

    \includegraphics[width=\linewidth]{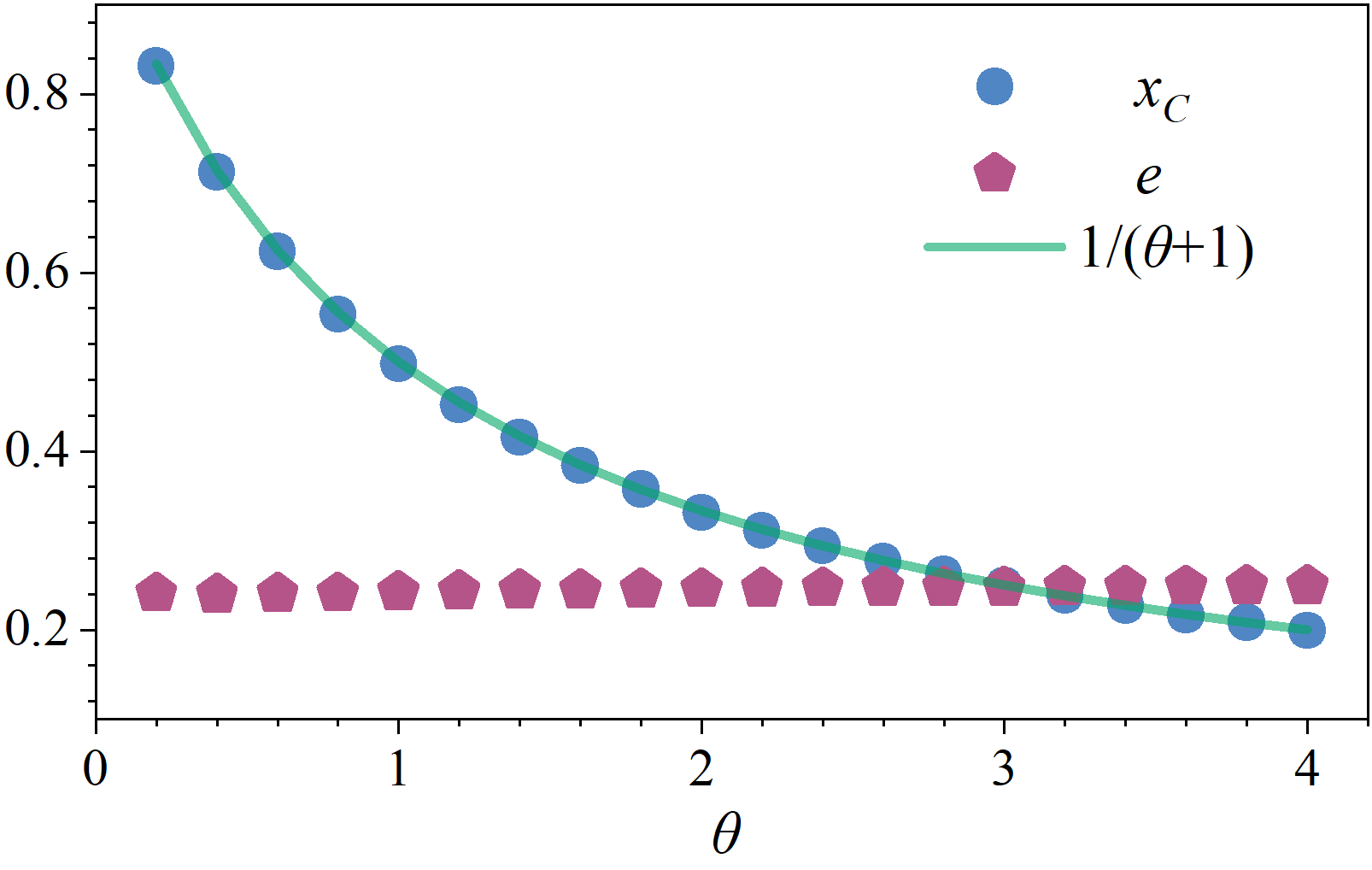}

  \caption{\label{fig6}Critical value $\mu_C$ of the transition from the oscillating state to the equilibrium state as a function of parameter $\theta$ of simulations, is represented by the dividing line between the yellow (oscillating state) and green (equilibrium state) regions. Each critical value $\mu_C$ is averaged over 50 repetitions. The orange dashed line indicates the critical value $\mu_C$ of numerical results, which do not exhibit the transition from the oscillating state to the equilibrium state for low values of $\theta$.}    
    
\end{figure}

Figs.~\ref{fig5} and \ref{fig6} further verify the previous conclusion that, by keeping the value of $\theta$ constant while increasing $\mu$, the system transitions from the oscillating state to the equilibrium state. 
In both numerical and simulation results, for different values of $\theta$, the systems exhibit persistent oscillations (represented by two or more widely separated points with $x_C$ ($e$) values that share the same $\mu$ value) at low values of $\mu$ and reach the equilibrium state (represented by two or more points with nearly identical $x_C$ ($e$) values and the same $\mu$ value) at sufficiently large values of $\mu$ (see Fig.~\ref{fig5}). 
Moreover, in the simulation results of Fig.~\ref{fig5}, we did not exclude systems that did not stabilize within the time range $t=35\thinspace 000$ to $t=40\thinspace 000$ because the time for the system to stabilize near the critical value of $\mu$ was unpredictable. 
Consequently, regions of highly unstable oscillating amplitudes are observed near the critical value of $\mu$ in Fig.~\ref{fig5}, demonstrating that fluctuation phenomena are pronounced. 

As shown in Fig.~\ref{fig6}, the critical value $\mu_C$ of the simulations is correlated with $\theta$. 
Each critical value $\mu_C$ is averaged over 50 repetitions. 
In the phase space of $\theta$ and $\mu$, the green region indicates the equilibrium state, while the yellow region signifies the oscillating state. 
Within the range of $(0, 5.0]$ (noting that $\theta = 0$ is meaningless), the critical value $\mu_C$ initially increases and then decreases as $\theta$ increases. 
The maximum $\mu_C$ occurs at $\theta = 1.0$, where cooperators produce resources at the same rate that defectors consume them. 
As $|\theta - 1.0|$ increases, the disparity between the rates of resource production by cooperators and consumption by defectors also widens, leading to a decrease in the critical value $\mu_C$. 
Thus, systems more easily enter the equilibrium state when individuals of one strategy have a speed advantage in influencing resources. 

The critical values of the numerical results (orange dashed line) show similar qualitative trends to the simulation results, though the quantitative critical values differ significantly (see Fig.~\ref{fig6}). 
As mentioned in Sec.~\ref{mft}, the equations derived using the well-mixed mean-field theory do not fully capture the resource heterogeneity of the system and the spatial properties of the square lattice. 
Moreover, the stochastic dynamics observed in MC simulations often reveal phenomena (e.g., self-organized pattern formation) that extend beyond the deterministic parameter settings predicted by the mean-field equations \cite{asllani2013linear}. 
Comparing the numerical results reveals that the emergence of large-size clustered patterns near the critical value of $\mu$ significantly enhances system stability, resulting in lower values of $\mu_C$ in simulations. 
This effect of spatial distribution cannot be accurately described by Eqs.~\eqref{abx} and \eqref{abe}. 
However, these equations qualitatively describe the phenomenon of dynamic stability changes in the system as $\mu$ and $\theta$ vary. 
The system transitions from unstable to stable as $\mu$ increases, and the rate of convergence further accelerates after stabilization, which aligns with the simulation results (see Fig.~\ref{fig3}). 
If we rely solely on Replicator Dynamics to describe our model without incorporating resource heterogeneity, we arrive at the same equations as those presented by Weitz \textit{et al.}~\cite{weitz2016oscillating}, which fails to accurately capture the aforementioned phenomena.

Compared to the works of Weitz \textit{et al.}~and Tilman \textit{et al.}~\cite{weitz2016oscillating,tilman2020Evolutionary}, our model also demonstrates oscillating dynamics of the fraction of cooperators and the average resources of the system. 
The system cycles orderly through the four previously mentioned states (see Fig.~\ref{fig1} and Sec.~\ref{rd_a}). 
However, their studies employed replicator dynamics to investigate the coevolutionary game dynamics with environmental feedback and ignored the influence of spatial heterogeneity. 
The transitions from oscillating to stable states in their research were directly induced by changes in dynamic stability due to payoff matrices and other parameters. 
In contrast, the transition from the oscillating state to the equilibrium state in our work is driven by dynamic stability, taking spatial heterogeneity into account, and can be influenced by self-organized spatial clustered patterns of cooperators, defectors, and resources (see results and discussions of Fig.~\ref{fig6}). 
The work of Lin \textit{et al.}~\cite{lin2019Spatial} also considered spatial heterogeneity in environmental resources, but their focus was primarily on the impact of payoff matrices and the diffusion of environmental resources on stability of the system. 
In contrast, our study does not account for the diffusion of resources. 
Moreover, we focus on investigating the influence of the resource update rate (i.e., $\theta$ and $\mu$) and the spatial distributions of strategies and resources on the emergence of the o-TOC. 

Hassell \textit{et al.}~found that lattice structures, spiral waves, and spatial chaos patterns effectively maintain population persistence in the host-parasitoid system under conditions of spatial heterogeneity and local movement patterns \cite{hassell1991Spatiala}. 
Although our work does not exhibit identical phenomena, both our findings and theirs illustrate similar self-organized criticality in complex systems with spatial heterogeneity. 
Besides, previous studies of real-world ecosystems have also demonstrated relations between spatial patterns and system stability \cite{liu2014pattern, rietkerk2021evasion, siteur2014beyond, bastiaansen2018multistability, bastiaansen2020effect}. 
Our results confirm that self-organized clustering behavior in coevolutionary game systems with spatially heterogeneous structures can also impact system stability.

\begin{figure}

    \includegraphics[width=\linewidth]{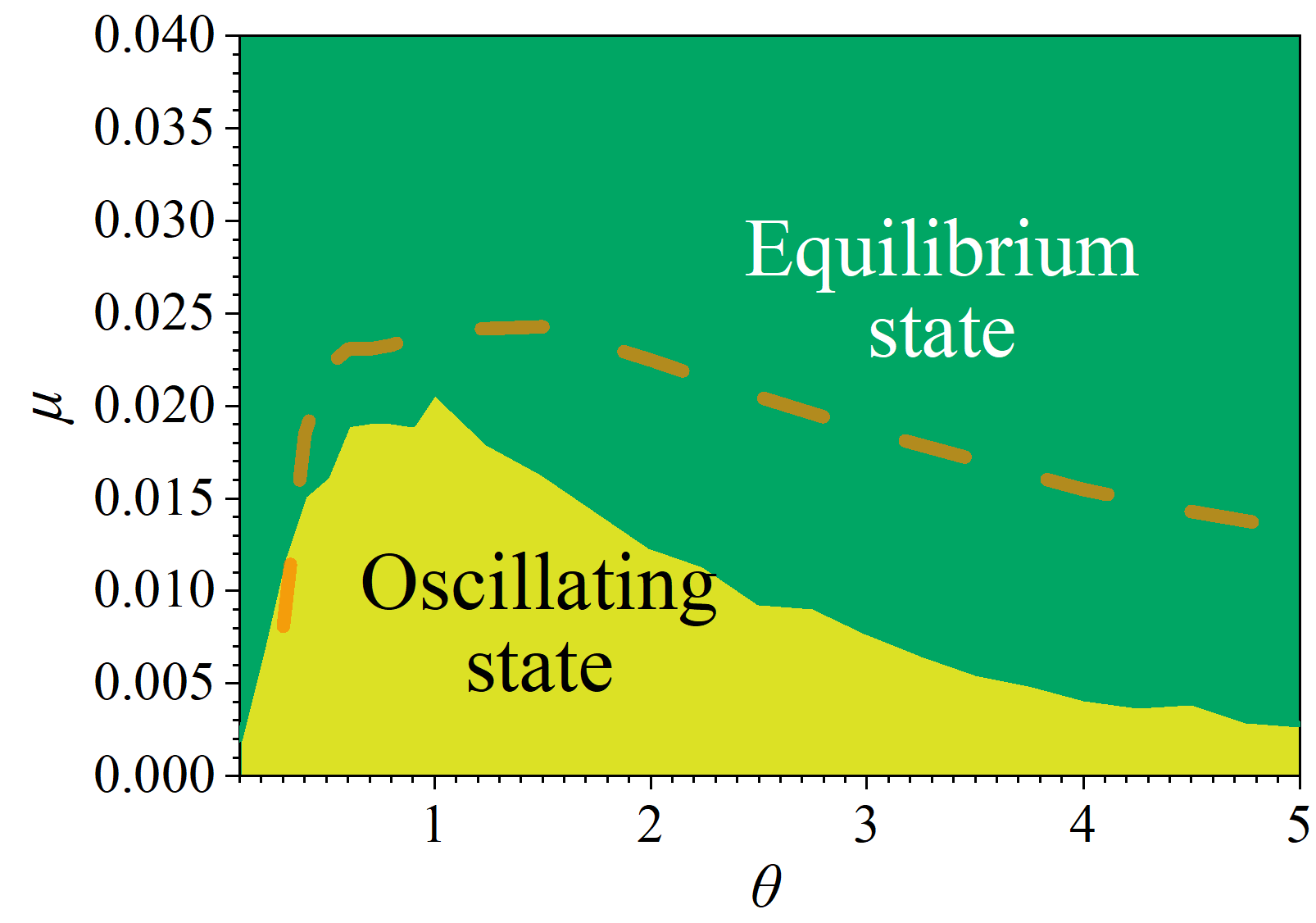}
  \caption{\label{fig7}The fraction of cooperators and the average resources of the system as functions of $\theta$ for $\mu=0.002$. Each point of $x_C$ or $e$ is averaged over ten repeated simulations. The green line represents the theoretical approximate value of the fraction of cooperators calculated by Eq.~\eqref{eq_de2}. }
    
\end{figure}

For further analysis, we observed the average values of the fraction of cooperators and the average resources of the system across different $\theta$ values (see Fig.~\ref{fig7}). 
The fraction of cooperators $x_C$ decreases rapidly as $\theta$ increases, while the average resources of the system $e$ remains relatively unchanged. 
This result suggests that individuals in the system tend to maintain environmental resources at an approximately fixed level, regardless of the proportion of cooperators and defectors, due to the parameter $\theta$. 
Thus, there is a more straightforward method for calculating the fixed value of the fraction of cooperators compared with solving Eqs.~\eqref{abx} and \eqref{abe}. 
In our model, defectors consistently consume resources at a constant rate whereas cooperators produce resources at a rate proportional to $\theta$. 
Consequently, the average variation of resources per MC step can be expressed as: 
\begin{equation}\label{eq_de1}
    \Delta e = \mu (\theta x_C - (1 - x_C)).
\end{equation}
Setting $\Delta e = 0$ yields the fixed point for the fraction of cooperators:
\begin{equation}\label{eq_de2}
    x_C^{*} = \frac{1}{1 + \theta}.
\end{equation}
This theoretical prediction aligns well with the simulation results and offers a more straightforward approach to determine $x_C^*$ compared with solving Eqs.~\eqref{abx} and \eqref{abe}. 
However, a more accurate characterization of the complete dynamics of the system still requires the use of Eqs.~\eqref{abx} and \eqref{abe}. 
Specifically, all the points representing the fraction of cooperators in the simulations fall on the curve given by $x_C^{*} = 1/(1 + \theta)$ (see Fig.~\ref{fig7}).

Based on the results in Fig.~\ref{fig7} and Eq.~\eqref{eq_de2}, the parameter $\theta$ influences the average fraction of cooperators, with $\theta = 1.0$ approximately representing the condition of an equal number of cooperators and defectors. 
Therefore, cooperators are in the minority when $\theta > 1.0$ and become the majority when $\theta < 1.0$. 
Additionally, the critical value $\mu_C$ for the transition is correlated with $|\theta - 1.0|$ (see Fig.~\ref{fig6}). 
Consequently, the fraction of cooperators is also a potential influencing factor for the critical value $\mu_C$ in simulations. 
Besides, the initial distributions of strategies and resources (e.g. cooperators are initially clustered) can impact this critical value, which we will explore in future studies.

\begin{figure}
    \includegraphics[width=\linewidth]{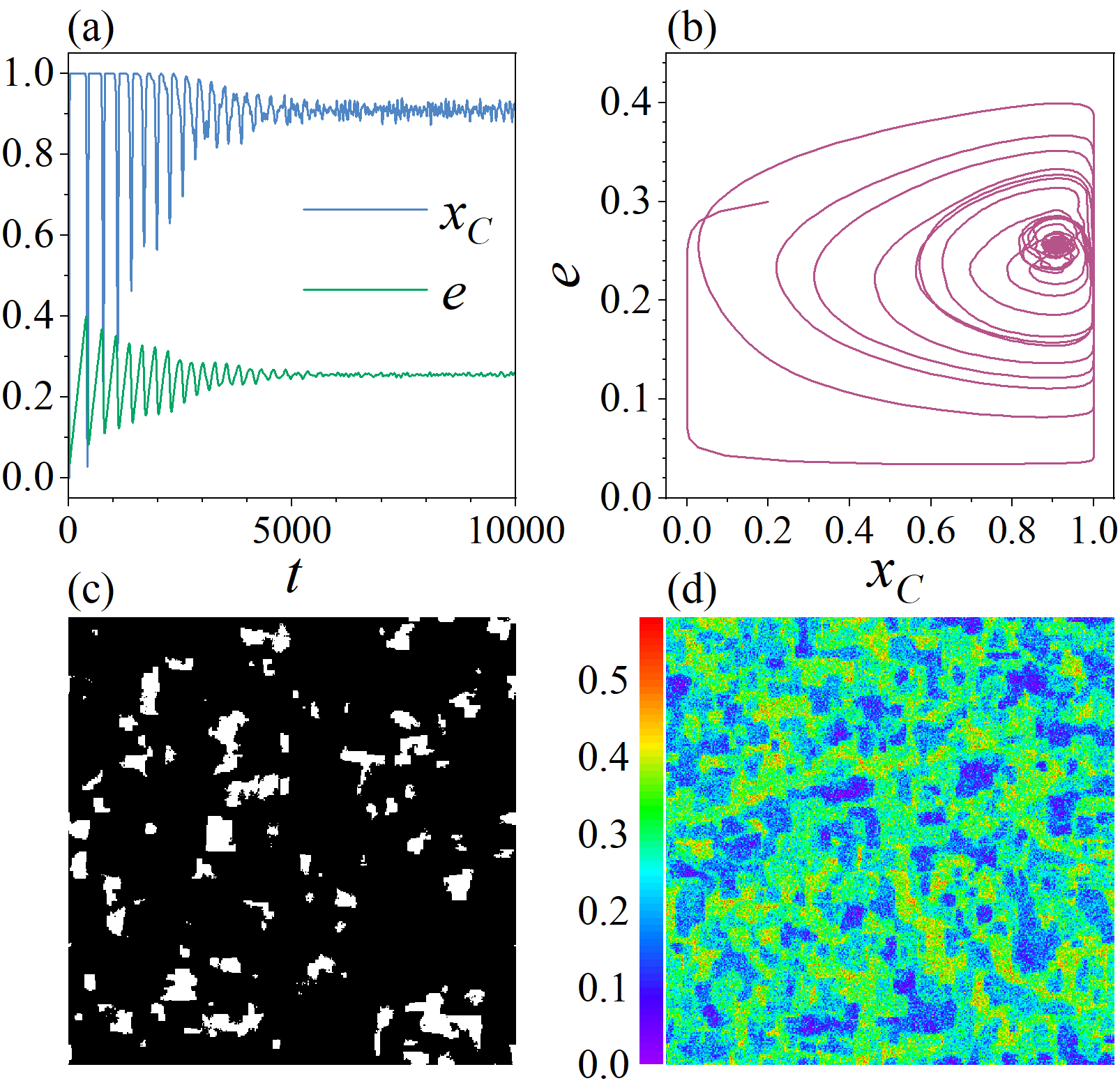}

  \caption{\label{fig8}Equilibrium state and irregular block clusters emerge for $\theta=0.1$, $\mu=0.01$. (a) Time evolutions of fraction of cooperators and average resources of the system. (b) Phase diagram of the coevolutionary dynamics of strategies and resources. (c) Distribution of cooperators (black pixels) and defectors (white pixels). (d) Distribution of environmental resources of nodes in the system. }    
  
\end{figure}

Moreover, when $\theta$ decreases sufficiently close to $0.1$ or lower (i.e., cooperators produce resources at a significantly slower rate than defectors consume them), and $\mu$ is sufficiently large, irregular block clusters form [see Figs.~\ref{fig8}(c) and (d)], and the transition from the oscillating state to the equilibrium state occurs [see Figs.~\ref{fig8}(a) and (b)]. 
It is crucial to reiterate that a low value of $\theta$ results in a high fraction of cooperators [see Eq.~\eqref{eq_de2}]. 
In this scenario, cooperators occupy most of the system's space, while defectors gather in irregular block clusters, as shown in Fig.~\ref{fig8}(a). 
Figure \ref{fig8}(b) illustrates that nodes with similar resource levels also cluster in irregular blocks. 
During the formation of these clusters, no spiral patterns are observed; instead, block clusters appear sequentially and gradually occupy the system. 
However, in the numerical results, the transition from the oscillating state to the equilibrium state is not observed when the value of $\theta$ is low (see Fig.~\ref{fig6}), highlighting the influence of spatial heterogeneity within the square lattice structure used in the simulations.

\section{\label{conclusion}Conclusion}

In this study, we investigate the coevolution of individuals' strategies and localized environmental resources in the prisoner's dilemma game model. 
Cooperators produce resources while defectors consume them. 
Individuals update their strategies based on myopic rules \cite{sysi2005spatial,chen2009cooperation,roca2009promotion}, which are solely correlated with instantaneous payoffs determined by the resources in their vicinity and the strategies of their neighbors. 
The o-TOC \cite{weitz2016oscillating,lin2019Spatial,tilman2020Evolutionary} manifests when the update rate of resources $\mu$ is sufficiently low, the system cycles orderly through four states:($x_{C,\mathrm{min}}$, $e_{\mathrm{min}}$), ($x_{C,\mathrm{max}}$, $e_{\mathrm{min}}$), ($x_{C,\mathrm{max}}$, $e_{\mathrm{max}}$), and ($x_{C,\mathrm{min}}$, $e_{\mathrm{max}}$). 
Moreover, a transition from the oscillating state to the equilibrium state arises when $\mu$ exceeds the critical value $\mu_C$ in the simulation results. 
We demonstrate that the critical value is primarily governed by the strength of cooperators in restoring the environment, denoted as $\theta$, and can be affected by system size $N$, spatial structure, noise $k$ and the initial distributions of cooperators. 
When individuals sharing the same strategy exert a greater influence on resources than others, it becomes easier for them to maintain system stability.  
Additionally, we employ mean-field theory and numerical solutions to obtain a theoretical approximation of the temporal evolution of the fraction of cooperators and the average resources of the system in a well-mixed structure, considering environmental heterogeneity. 
In this context, we observe the same qualitative trends regarding the aforementioned transition and greater critical values of $\mu$ in the numerical results. 
Further findings indicate that self-organized clustered patterns of simulations on a square lattice significantly enhance the stability of the system for $\mu$ values near $\mu_C$. 
This effect is not captured in the numerical results, where individuals randomly choose neighbors for single interactions in a well-mixed structure, resulting in a lack of stable distributions. 
Similarly, previous studies of real-world ecosystems have shown that systems with specific spatial structures can mitigate catastrophic instability outcomes \cite{liu2014pattern, rietkerk2021evasion, siteur2014beyond, bastiaansen2018multistability, bastiaansen2020effect}. 
Overall, we demonstrate that environmental heterogeneity and stable structures (which can lead to stable distributions) contribute to individuals' ability to avoid the o-TOC in coevolutionary game systems. 
In future work, we intend to further explore the influences of initial distributions, payoff matrices, and other factors not discussed here.

\appendix*

\section{\label{app_a}Individual's probability of cooperation}
\setcounter{figure}{0}
\renewcommand{\thefigure}{A\arabic{figure}}

\begin{figure}
    \includegraphics[width=\linewidth]{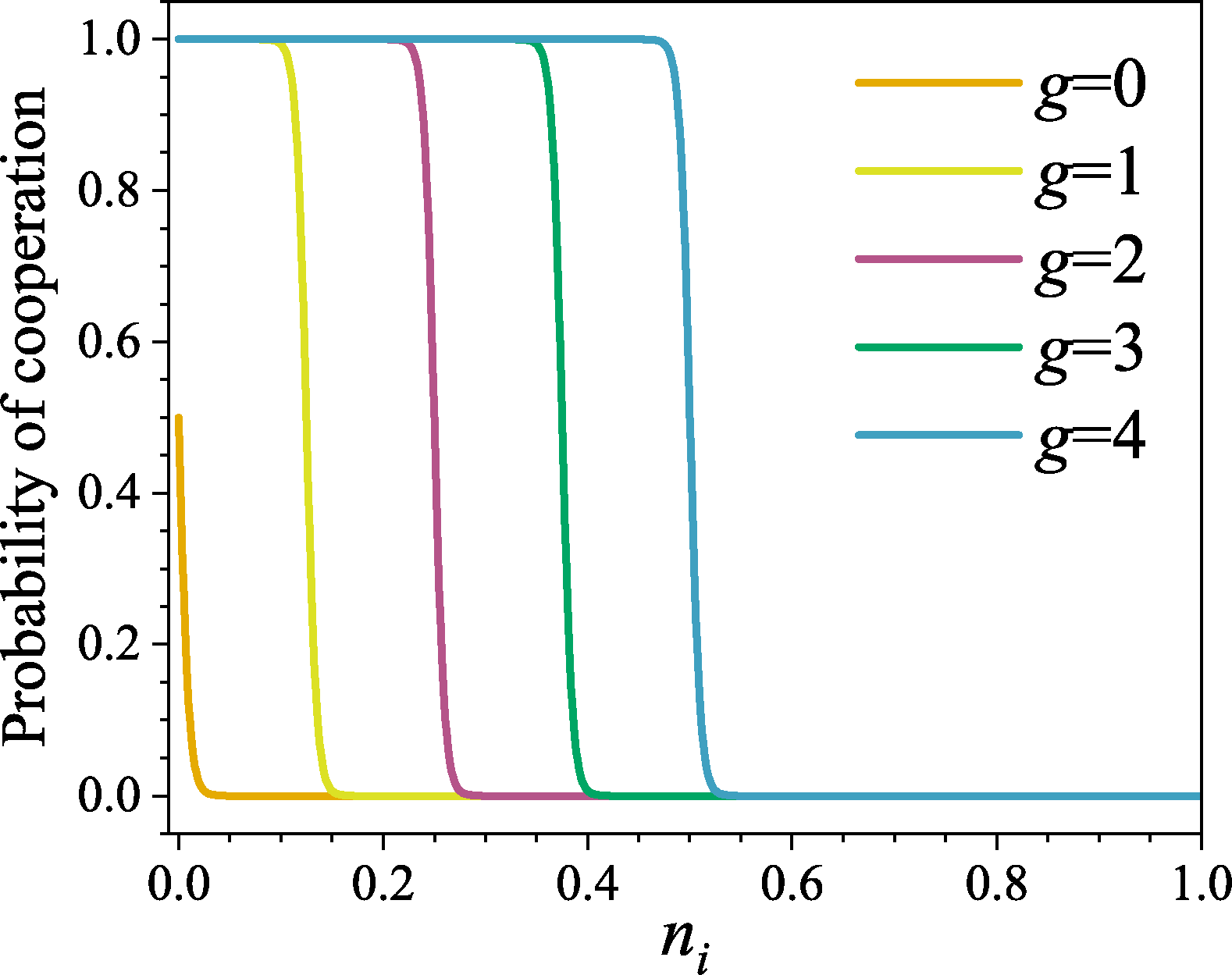}
  \caption{\label{figa1}Individual's probability of cooperation as a function of usable resources $n_i$ for different numbers of cooperative neighbors $g=0, 1, 2, 3, 4$. }    
\end{figure}

Each individual's payoff is determined by its own and its neighbors' strategies and the payoff matrices. 
Based on Eqs.~\eqref{e2} and \eqref{e3}, the payoff matrices are coupled with individual's usable resources, thus, any individual $i$'s average payoff is:
    \begin{equation}\label{aa1}
        \begin{aligned}
            \pi_i=
            \begin{cases}
                &0.25\times[(1-n_i)\times(gR_0+(4-g)S_0)\\
                &+n_i\times(gR_1+(4-g)S_1)] \qquad s_i=C, \\
                ~\\
                &0.25\times[(1-n_i)\times(gT_0+(4-g)P_0)\\
                &+n_i\times(gT_1+(4-g)P_1)] \qquad s_i=D, \\
            \end{cases}
        \end{aligned}
    \end{equation}
where $g=0, 1, 2, 3, 4$ is the number of cooperators among individual $i$'s neighbors. 
According to Eq.~\eqref{rpf}, the payoffs in Eq.~\eqref{aa1} can be calculated as:
    \begin{equation}\label{aa2}
        \begin{aligned}
            \pi_i=
            \begin{cases}
                &0.25\times[(1-n_i)\times(5g+2(4-g))\\
                &+n_i\times(5g+0(4-g))]\\
                &=0.25(3g+2n_i g-8n_i+8) \qquad s_i=C,\\
                ~\\
                &0.25\times[(1-n_i)\times(4g+2(4-g))\\
                &+n_i\times(6g+2(4-g))]\\
                &=0.25(2g+2n_i g+8) \qquad s_i=D.\\
            \end{cases}
        \end{aligned}
    \end{equation}
Thus, the difference in payoffs for individual $i$ between cooperation and defection is:
    \begin{equation}\label{aa3}
        \begin{aligned}
            \Delta\pi_i&=\pi_{i(C)}-\pi_{i(D)}\\
            &=0.25(3g+2n_i g-8n_i+8)-0.25(2g+2n_i g+8)\\
            &=0.25g-2n_i,
        \end{aligned}
    \end{equation} 
where $\pi_{i(C)}$ is individual $i$'s payoff when $s_i=C$ and $\pi_{i(D)}$ is individual $i$'s payoff when $s_i=D$.

If individual $i$'s original strategy is cooperation, its probability of choosing defection is:
    \begin{equation}\label{aa4}
        \begin{aligned}
            P_{s_i}(D \rightarrow C)&=\frac{1}{1+{\rm exp}((\pi_{i(C)}-\pi_{i(D)})/k)}\\
            &=\frac{1}{1+{\rm exp}(\Delta\pi_i/k)}. 
        \end{aligned}
    \end{equation}
Thus, its probability of remaining cooperation is:
    \begin{equation}\label{aa5}
        \begin{aligned}
            P_{s_i}(C \rightarrow C)&=1-P_{s_i}(D \rightarrow C)\\
            &=1-\frac{1}{1+{\rm exp}(\Delta\pi_i/k)}. 
        \end{aligned}
    \end{equation}
And if individual $i$'s original strategy is defection, its probability of choosing cooperation is:
    \begin{equation}\label{aa6}
        \begin{aligned}
            P_{s_i}(C \rightarrow D)&=\frac{1}{1+{\rm exp}((\pi_{i(D)}-\pi_{i(C)})/k)}\\
            &=\frac{1}{1+{\rm exp}(-\Delta\pi_i/k)}. 
        \end{aligned}
    \end{equation}
It is easy to prove that $\frac{1}{1+{\rm exp}(x)} + \frac{1}{1+{\rm exp}(-x)} = 1$. 
Based on Eqs.~\eqref{aa3}, \eqref{aa5} and \eqref{aa6}, we have:
    \begin{equation}\label{aa7}
        \begin{aligned}
            &P_{s_i}(C \rightarrow D)=\frac{1}{1+{\rm exp}(-\Delta\pi_i/k)}\\
            &=1-\frac{1}{1+{\rm exp}(\Delta\pi_i/k)}=P_{s_i}(C \rightarrow C)\\
            &=\frac{1}{1+{\rm exp}((2n_i-0.25g)/k)}. 
        \end{aligned}
    \end{equation}
Thus, regardless of any individual's previous strategy, the probability that it updates to cooperative strategy is the result shown in Eq.~\eqref{aa7}. 
And any individual's probability of choosing defection is also independent of its previous strategy:
    \begin{equation}\label{aa8}
        \begin{aligned}
            &P_{s_i}(D \rightarrow D)=1-P_{s_i}(C \rightarrow D)\\
            &=1-P_{s_i}(C \rightarrow C)=P_{s_i}(D \rightarrow C)\\
            &=\frac{1}{1+{\rm exp}((0.25g-2n_i)/k)}. 
        \end{aligned}
    \end{equation}

The results in Fig.~\ref{figa1} can be calculated from Eq.~\eqref{aa7}. 
The probability of choosing cooperation decreases as an individual's available resources increase. 
Furthermore, an individual's probability of cooperation diminishes rapidly to nearly zero when the average resources of the system approach a critical value determined by the number of cooperative neighbors $g$. 
The greater the number of cooperators in the neighborhood, the higher the critical value.


\begin{acknowledgments}
    The authors warmly thank S.-J.~Wang at Shaanxi Normal University  for helpful discussions. 
    This work was supported by the National Natural Science Foundation of China (Grants Nos.~11475074, 12247101 and 12375032), the Fundamental Research Funds for the Central Universities (Grant Nos.~lzujbky-2024-11, lzujbky-2023-ey02, lzujbky-2019-85 and lzujbky-2024-jdzx06), the Natural Science Foundation of Gansu Province (No.~22JR5RA389), and the 111 Center under Grant No.~B20063. 
    This work was partly supported by Longyuan-Youth-Talent Project of Gansu Province. 
    The authors extend their gratitude to the editors and referees for their diligent efforts during the submission process and invaluable suggestions, which have significantly enhanced this paper.
\end{acknowledgments}

\bibliography{refle}

\end{document}